\documentclass[jkps,preprint,fleqn,showpacs,showkeys]{revtex4}
\usepackage{graphicx}
\usepackage{amssymb}
\usepackage{amsmath}
\usepackage{bm}
\begin{document}
\setcounter{page}{0}
\title[]{Magnetism on the boron-doped Si(111)-$\sqrt 3 \times \sqrt 3$ surface}
\author{Chang-Youn \surname{Moon}}
\email{cymoon@kriss.re.kr}
\author{Daejin \surname{Eom}}
\author{Ja-Yong \surname{Koo}}
\affiliation{Quantum technology institute, Korea Research Institute of Standards
and Science, Yuseong, Daejon 34113, Republic of Korea}

\date[]{}

\begin{abstract}
We investigate the possible magnetism on the Si(111)-$\sqrt 3 \times \sqrt 3$
surface, which is stabilized for highly boron-doped samples, by using
first-principles calculations. When the silicon adatom on top of a boron
atom is removed to form a defect structure, three silicon dangling bonds are
exposed generating half-filled doubly degenerate energy levels in the band gap
which stabilizes a local magnetic moment of 2 $\mu_{\rm B}$. When there are many
such defect structures adjacent one another, they are found to align
antiferromagnetically. However,
we demonstrate that the ferromagnetism can be stabilized by adjusting the
number of electrons in defects, suggesting a possibility towards
spintronic applications of this unique silicon surface structure.
\end{abstract}



\maketitle

\section{INTRODUCTION}
Exploiting the spin as well as charge degrees of freedom of materials
in information processing, nowadays often referred to as spintronics,
composes one of the major parts of modern day condensed matter physics
and materials science, as it is considered as a promising candidate
to overcome the limiting factors of conventional semiconductor technologies
based on utilizing only the charge degree of freedom. Materials used
for this purpose should be magnetically ordered, in most cases ferromagnetically,
which can produce the spin-polarized current. The majority of efforts to
obtain a magnetically ordered state has been made using conventional
magnetic materials containing 3$d$-active elements such as Mn, Fe, Co,
etc., as the strong intra-atomic Coulomb interaction within localized 3$d$
orbitals tends to induce the local magnetic moment. In these materials,
the magnetic elements are periodically ordered to form a crystal or at
least incorporated as impurities as in the case of diluted magnetic
semiconductors, mostly with compound semiconductors such as ZnO,
TiO$_2$, and GaAs, etc., as host \cite{DMS1,DMS2,DMS3}.

Meanwhile, magnetism originating purely from atomic $s$ and/or $p$
orbitals has not been reported very often, with past few studies
mostly based on the carbon-based nanostructures \cite{yhkim,ywson1,ywson2}.
If a magnetic ordering can be realized in a technologically more mature
and versatile material such as silicon, spintronic applications would
develop much faster and in more diverse ways.
Authors of the present paper has been studying peculiar structural
and electronic properties of the Si(111)-$\sqrt 3 \times \sqrt 3$ surface, which
is stabilized for highly boron-doped samples \cite{Si111_1}. In this surface, a top silicon atom,
or silicon adatom (Si$_{\rm ad}$), is triply coordinated with other silicon atoms
underneath which in turn are bonded to a boron atom below \cite{Si111_2} (see Fig. 1).
It was found that the position of Si$_{\rm ad}$ is the key to switching between two
very distinct quantum states, with which the possible application to a very
high-density memory device was demonstrated using the scanning tunneling
microscopy (STM) technique \cite{EMK2015}. Regarding Si$_{\rm ad}$, further
interesting question would be what happens when we remove a Si$_{\rm ad}$
leaving three adjacent partially-filled dangling bonds
of underneath silicon atoms, which seems promising for holding a large
local magnetic moment.

In the present work, we perform a theoretical study of the magnetic properties
of boron-doped Si(111)-$\sqrt 3 \times \sqrt 3$ surface using first-principles
calculations based on the density-functional theory. Si$_{\rm ad}$ vacancy (V${_{\rm Si}}_{\rm ad}$),
a defect structure with the missing Si$_{\rm ad}$, has three dangling-bond orbitals
from each of the three silicon atoms, which are filled with two electrons in total.
Spin-polarized calculation reveals that a net magnetic moment of 2 $\mu_{\rm B}$,
which is equally distributed over the three dangling-bond orbitals, is stabilized
for an isolated V${_{\rm Si}}_{\rm ad}$. When there are series of the defect structures
adjacent with one another, the anti-parallel arrangement of magnetic moments turns out
to be more stable than the parallel alignment, as expected for the exchange interaction
among half-filled orbitals. We further investigate the doping dependence of the
exchange interaction between the defects, and find that the hole doping enhances
the relative stability of the ferromagnetic (FM) ordering with respect to that of the
antiferromagnetic (AFM) ordering, and over 0.1 hole per defect eventually the FM state
becomes more stable than the AFM one. Our result suggests a new possibility for
the spintronic application of silicon, the most well understood and widely used
conventional semiconductor.

\section{Computational detail}
Our first principles calculations are based on the density functional theory with
the PAW potential \cite{PAW1,PAW2} as implemented in the VASP code \cite{VASP1,VASP2}.
Electronic wave functions are expanded with plane waves up to a cutoff energy of 318 eV.
A slab structure of 10 silicon monolayer thick (16 \AA) which consists of $2\times2$ primitive 
units of the $\sqrt{3} \times \sqrt 3$-surface structure including 4 boron atoms, is
constructed to mimic
the Si (111) surface, where the dangling bonds on the bottom silicon layer are passivated
by hydrogen atoms. A vacuum
region of about 7 \AA\ separates the slab and its periodic images along the surface normal
direction. We adopt the experimental lattice constant in determining the in-plane supercell
dimensions, and all the surface structures studied in this work are fully relaxed
both for spin-unpolarized and polarized calculations if not indicated otherwise, except for
the silicon atoms on the bottom layer representing the rigidity of the bulk region.
The K-point sampling is performed on a $8\times8\times1$ Monkhorst-Pack grid \cite{MP}.

\section{Result and discussion}
Fig. 1 depicts the atomic structure of Si(111)-$\sqrt 3 \times \sqrt 3$
surface, which is stable for highly boron-doped
silicon \cite{Si111_1}. Every boron atom, substitutionally located on the
third atomic layer, is accompanied by a Si$_{\rm ad}$ sitting directly
above. A hole generated by the substitutional boron
compensates an electron from the lone-pair orbital of Si$_{\rm ad}$,
leaving the orbital empty and making the surface structure energetically stable
and chemically inert. If a Si$_{\rm ad}$ is removed to form V${_{\rm Si}}_{\rm ad}$
by a suitable
mean, e.g., the STM tip manipulation \cite{EMK2015}, three silicon
atoms on the second atomic layer are exposed to the vacuum with a dangling-bond orbital
of each atom, pointing up perpendicular to the surface plane as illustrated in
Fig. 1(b).

With the $C_{3v}$ symmetry, the three dangling-bond orbitals in
V${_{\rm Si}}_{\rm ad}$ form {\it two-fold degenerate} molecular orbital ($E$)
energy levels inside the band gap of silicon.
As one electron is captured by an
acceptor state generated by the nearby boron, two electrons are
left for the molecular orbitals and hence the $E$ orbitals are
half-filled, as represented by a peak structure across the
Fermi level ($E_{\rm F}$) in the calculated spin-unpolarized density of
states (DOS) in Fig. 2(b). Since two electrons occupy the doubly degenerate
orbitals which are spatially localized, the Hund's coupling dictates
the total spin of electrons to be maximized, resulting in a triplet
state with a magnetic moment of 2 $\mu_{\rm B}$.
Indeed, our spin-polarized calculation leads to
the exchange splitting between spin-up and spin-down states inside
the band gap as shown in Fig. 2(b). The gain of the total energy
by this spin polarization is 187 meV, which lies in the typical
energy range of Hund's coupling. The calculated spin density,
which is defined as the difference between charge densities of
spin-up and spin-down states, is displayed in Fig. 2(a).
It clearly shows that the spin density profile is symmetrically distributed over the
three silicon atoms, with some additional weight
on the boron atom (in opposite sign) and other silicon atoms
nearby. Because of the missing Si$_{\rm ad}$,
the exposed three silicon atoms are pulled down from the original positions
with Si$_{\rm ad}$
and form nearly planar $sp^2$ bondings with the surrounding atoms.
As a consequence,
the spin density becomes almost atomic $p_z$-like. Therefore,
we conclude that a local magnetic moment of 2 $\mu_{\rm B}$ is formed
on a defect structure made by removing a Si$_{\rm ad}$.

Although the formation of a local magnetic moment in this single defect structure 
is fascinating itself and possibly can serve as a new platform for the
quantum spin device application, it is also of importance and interest
to understand how the spin moments would align when there are many of
them in such a case where rather an extended area of polarized 
spins is desirable.
To dig out this matter, we build V${_{\rm Si}}_{\rm ad}$ defects in a row 
adjacent one another, and perform spin-polarized calculations to get a FM
solution where all the magnetic moments are in the same direction
(Fig. 3(a))
as well as an AFM solution where spin directions
alternate from one site to the next (Fig. 3(b)). It turns out that the
AFM alignment is energetically more favorable than the FM alignment by 36 meV
per supercell containing 2 V${_{\rm Si}}_{\rm ad}$.
This can be understood as a result of direct exchange interactions
between the local magnetic moments in half-filled localized defect orbitals, 
where anti-parallel
spin alignment is required for the electrons to hop between neighboring
defect orbitals to gain in the kinetic energy while observing
the Pauli exclusion principle \cite{moon2014}. As typical for the direct exchange,
the interaction is very short-ranged and the AFM state is only
8 meV more stable than the FM state for V${_{\rm Si}}_{\rm ad}$'s at the next-nearest
neighbor distance, which is estimated from a calculation using a larger supercell
($4 \times 2$ units of $\sqrt 3 \times \sqrt 3$ surface structure).
Because the exchange splitting of defect states
in an isolated V${_{\rm Si}}_{\rm ad}$ results in fully occupied majority spin orbitals
and empty minority spin orbitals as we see in Fig. 2(b), the band structure
from the series of the localized V${_{\rm Si}}_{\rm ad}$ orbitals is found to have
a band gap.

For the spintronic application purpose, the FM alignment would be more
suitable because it can be a source of spin-polarized electrons. As demonstrated
above, the half-filled orbitals prefer the AFM exchange coupling and the electron localization
(band gap). On the other hand, the fractional filling of orbitals facilitates
electron delocalization (partially-filled bands)) and the FM state can be
stabilized to maximize the Hund's coupling energy among the hopping electrons,
while the kinetic energy gain is the same between FM and AFM alignments
because the electron hopping is allowed for both spin alignments in this
partially filled orbital case.
To test this possibility, we investigate the stability of spin alignments
as a function of the hole doping, considering that many extra boron
dopants other than those in the third layer of the $\sqrt 3 \times \sqrt 3$ surface
can exist near the surface in this highly
boron-doped system. The result is shown in Fig. 3(c). Here we consider
not only the fully relaxed structure for each spin alignment and doping
level, but also the lattice structure fixed to that from the spin-unpolarized
calculation with no doping, to separate the purely electronic exchange
interaction from the lattice relaxation effect. In the fixed lattice
case, the FM state is less stable than the AFM state by 52 meV
for zero doping. On the other hand, the relative stability of the FM state increases
with hole doping, and over around 0.25 hole per defect the FM state
becomes the ground state. The maximum stability of the FM state is found to
be 29 meV at around 0.9 hole per defect. This
doping level is close to 0.5 electron per orbital in
doubly degenerate majority-spin energy levels (see Fig. 2(b)), where the electron
itinerancy (delocalization)
is maximized. This is the best condition for the FM state to be stabilized
with a purely electronic origin such as the electron hopping and Hund's
coupling. Meanwhile,
the effect of lattice relaxation is to further stabilize the FM state,
indicated by the overall downward shift of the curve with the lattice
relaxation in Fig. 3(c). Consequently, a doping as small as 0.1 hole
per defect can drive the system into the FM state, and at around 0.6
hole per defect the maximum stability of 39 meV is reached. At
this doping level, the total magnetization of the system is reduced
to 1.4 $\mu_{\rm B}$ per defect, and the system is metallic with a finite DOS
for the majority spin channel at $E_{\rm F}$ as shown in Fig. 3(d).
In contrast, the DOS value at $E_{\rm F}$ for the minority spin channel is
very small, resulting in the spin polarization, defined as
$P(E)=\frac{\rho_\uparrow (E)-\rho_\downarrow (E)}{\rho_\uparrow (E)+
\rho_\downarrow (E)}$, estimated to be 73 \% for $E=E_{\rm F}$,
being close to the half-metallicity or 100 \% spin polarization
at $E_{\rm F}$.

Our results show that a FM cluster of V${_{\rm Si}}_{\rm ad}$
defects can be formed when the Si(111)-$\sqrt 3 \times \sqrt 3$
surface is doped with holes, whereas the AFM coupling among magnetic moments is 
stable for the intrinsic surface. The FM state can be stable up to $\sim$390 K ($\sim$ 39 meV)
with a net magnetic moment of 1.4 $\mu_{\rm B}$ per V${_{\rm Si}}_{\rm ad}$ site
at a doping level of 0.6 hole per defect. This amount of doping
level can be considered to be feasible when we take into account a
possibility of the existence of many extra boron atoms near the 
surface in this highly boron-doped surface, as mentioned earlier. Moreover,
the FM state is highly spin-polarized at $E_{\rm F}$, making this system ideal
for the spintronic applications such as a source of the spin-polarized
current in a nanometer-sized device. Noteworthy is that silicon is also suitable
as a channel material for the spin transport \cite{Min,Appelbaum,Dash}
with the long spin
relaxation time and large spin diffusion length due to a weak
spin-orbit coupling of this material. Thus, a FM defect cluster on a silicon
surface would be very promising for the comprehensive spin device applications.
Finally, the electron doping might
have a similar effect with the hole doping, that is, stabilizing the FM 
coupling of defects.
However, the electron doping would fill unoccupied minority spin
orbitals above $E_{\rm F}$ of the undoped system, which is close to bulk
conduction band minimum (CBM) in our DFT calculation (see Fig. 2(b)).
These doped electrons can be delocalized by spilling into the bulk conduction
band states instead of being localized inside the defect orbital,
and, therefore, the mechanism of doping-driven stabilization of the FM state
might not work. On the other hand, this scenario can possibly be an artifact 
of the DFT calculation because the energy position of CBM is underestimated
as usual for DFT calculations. An appropriate band gap correction scheme,
such as using hybrid exchange-correlation functionals \cite{hse0}, would
predict a reliable behavior of the spin alignment with the electron doping,
which we leave as a future work.

\section{CONCLUSIONS}

In conclusion, we investigate a possibility of magnetic activities on the
Si(111)-$\sqrt 3 \times \sqrt 3$ surface by using the first-principles
calculation. Removing an Si$_{\rm ad}$ generates doubly-degenerate local
defect states consisting of three silicon dangling-bond orbitals,
which holds a local magnetic moment of 2 $\mu_{\rm B}$. While these local
moments tend to align anti-parallel in the intrinsic surface, hole
doping is found to enhance the stability of the FM alignment. The
maximum relative stability of the FM state with respect to the AFM state
is 39 meV implying that the FM state is stable up to 390 K for the 0.6 hole doping
per defect, with a local magnetic moment of 1.4 $\mu_{\rm B}$. Moreover,
the FM state exhibits a high spin polarization value at $E_{\rm F}$, making the
system promising for the spintronic applications.

\begin{acknowledgments}
This research was supported by the Basic Science Research
Program through the National Research Foundation of Korea
(NRF) funded by the Ministry of Science and ICT (2016R1C1B1014715).
\end{acknowledgments}

\newpage

\begin{figure}
\includegraphics[width=7in]{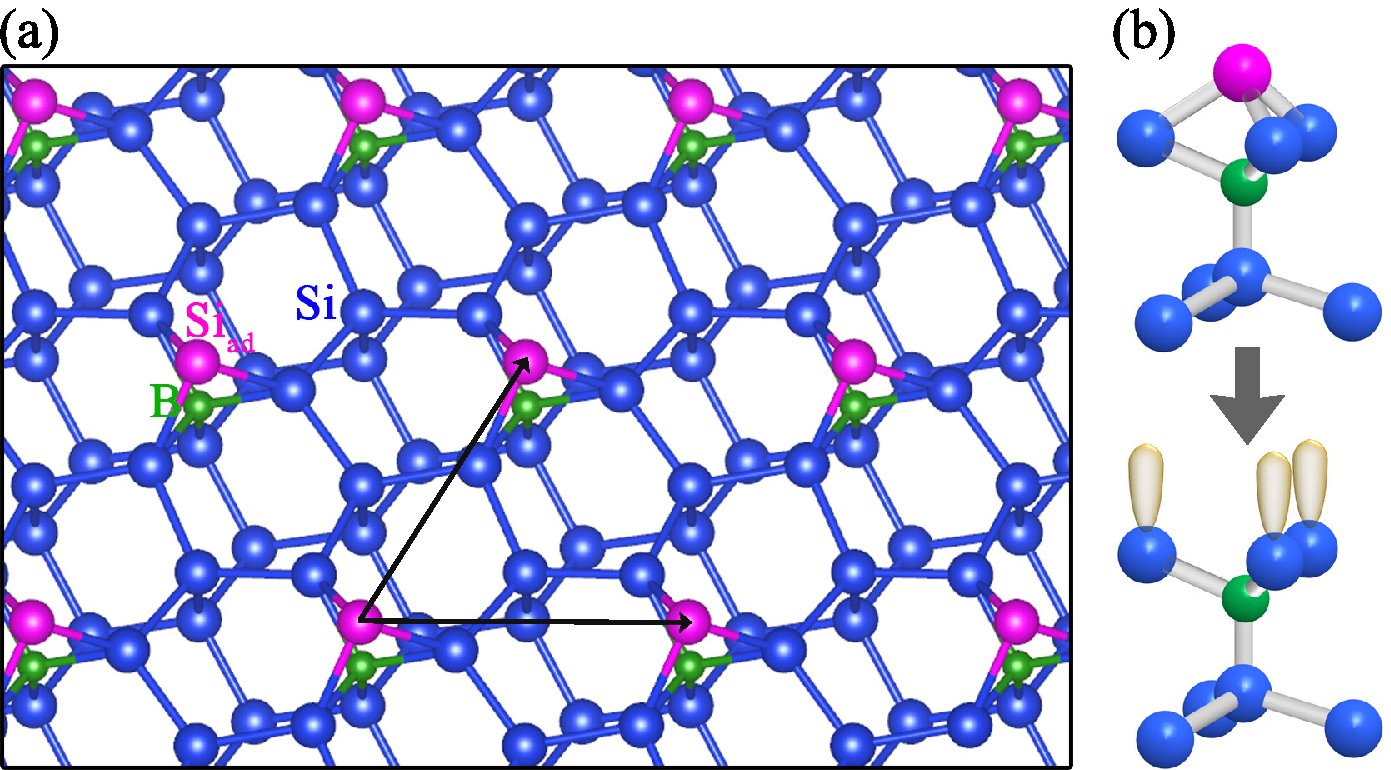}
\caption{(Color online) (a) Atomic structure of the Si(111)-$\sqrt 3
\times \sqrt 3$ surface. Only top few layers are shown for clarity. An
imaginary line joining a boron atom and the Si$_{\rm ad}$ atom on top of it defines
the surface normal direction, while in-plane primitive unit vectors are shown with
arrows. The supercell geometry used in this study contains $2 \times 2$
units of the in-plane primitive cell. (b) Schematics depicting the perfect
surface atomic arrangement (upper structure) and V${_{\rm Si}}_{\rm ad}$ (lower
structure) after removing a Si$_{\rm ad}$, exposing the silicon dangling-bond
orbitals. Note that Si$_{\rm ad}$ is represented in a color different from
that of other silicon atoms for clarity, although both are the same
silicon species.}
\label{fig1}
\end{figure}

\newpage

\begin{figure}
\includegraphics[width=7in]{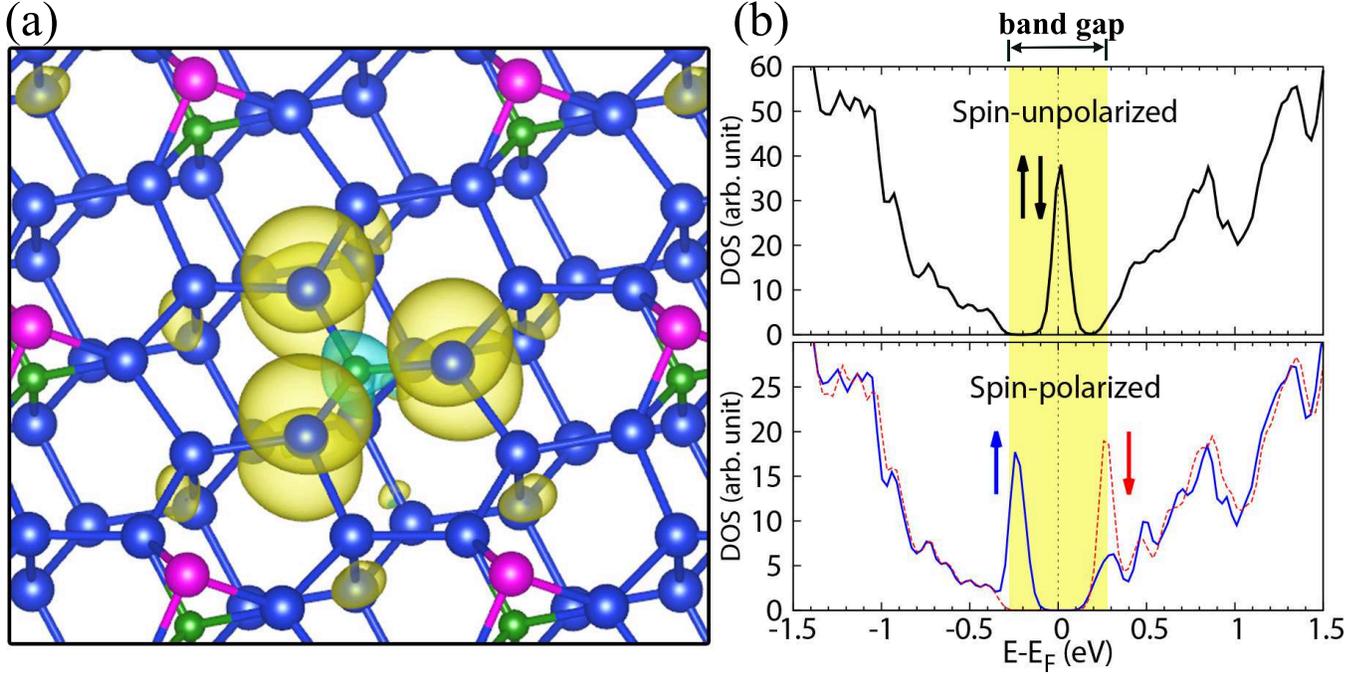}
\caption{(Color online)(a) Isosurface plot of the calculated spin
density in an isolated V${_{\rm Si}}_{\rm ad}$ defect. Yellow isosurface represents
the density of the majority spin state, while cyan color is for
the minority spin density represented by negative values, with absolute
value of 0.0013 $e/\AA^3$ for both spin density isosurface. (b) Total DOS of
the surface containing an isolated V${_{\rm Si}}_{\rm ad}$ defect. The upper
panel is from the spin-unpolarized calculation with doubly-degenerate
defect levels inside the band gap denoted by a shaded area, and the lower
panel from the spin-polarized calculation with majority and minority spin
defect levels split by an exchange energy.}
\label{fig2}
\end{figure}

\newpage

\begin{figure}
\includegraphics[width=7in]{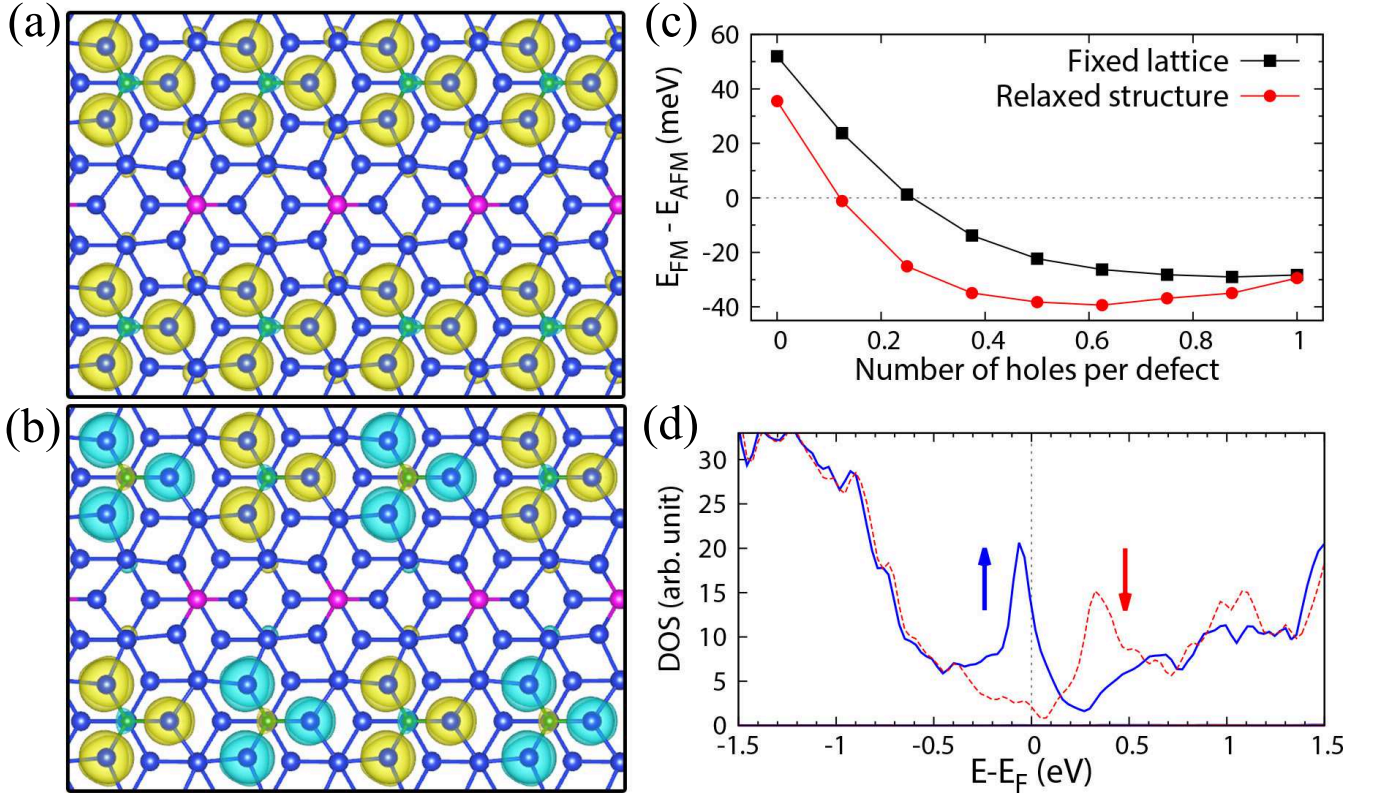}
\caption{(Color online) Isosurface plot of the calculated spin density
showing (a) the FM alignment (b) the AFM alignment of local defect
magnetic moments. (c) Total energy difference between the FM and AFM
states per supercell containing 2 V${_{\rm Si}}_{\rm ad}$ defects. For the
fixed lattice case, all the calculations are done using the same atomic
structure obtained by fully relaxing the 2-V${_{\rm Si}}_{\rm ad}$-defect
structure in a spin-unpolarized calculation, while in the relaxed structure
case the full structure optimization is taken for each spin alignment
and the doping level. Negative value of the energy difference represents
that the FM state is more stable than the AFM state. (d) Spin-resolved
DOS of the FM state for the doping level of 0.6 hole per defect. At $E_{\rm F}$,
DOS values for the opposite spin directions differ considerably, resulting
in a large value of spin polarization of 73 \% (see the text).}
\label{fig3}
\end{figure}

\end{document}